\documentclass[a4paper,12pt]{article}
 
\pdfoutput=1
\usepackage{jheppub}
\usepackage{dcolumn}
\usepackage{soul}
\usepackage[english]{babel}
\usepackage[utf8]{inputenc}
\usepackage{amsmath}
\usepackage{dcolumn}
\usepackage{soul}
\usepackage{hyperref}
\usepackage{url}
\usepackage{graphicx}
\usepackage{bm,amsmath,amssymb}
\usepackage[mathscr]{eucal}
\usepackage{makeidx}
\usepackage{subfig}
\usepackage{amsmath}
\usepackage{amssymb}
\usepackage{amsthm}
\usepackage{mathrsfs}
\usepackage{graphicx}
\usepackage{fancyhdr}
\usepackage{array}
\usepackage{simplewick}
\usepackage{latexsym}
\usepackage[all]{xy}
\usepackage{enumerate}
\usepackage{dsfont}
\usepackage{slashed}
\usepackage{float}
\usepackage[titletoc]{appendix}
\usepackage{ulem}
\usepackage{physics} 
\usepackage{tcolorbox} 

\usepackage{verbatim}

\usepackage{tikz,tikz-3dplot}
\tdplotsetmaincoords{80}{45}

\usepackage{tikz}
\usepackage{amsmath}
\usetikzlibrary{shapes.geometric, arrows.meta, positioning, fit, backgrounds}

\tikzset{surface/.style={draw=black, fill=white, fill opacity=.6}}

\usepackage{tikz}
\usetikzlibrary{decorations.pathmorphing,patterns}

\newcommand{\be}{\begin{equation}}
\newcommand{\ee}{\end{equation}}
\newcommand{\bea}{\begin{eqnarray}}
\newcommand{\eea}{\end{eqnarray}}

\newcommand{\ben}{\begin{eqnarray}}
\newcommand{\een}{\end{eqnarray}}

\title{\boldmath\LARGE{{Holographic Complexity as a Probe of Boundary Entropy in AdS/BCFT}}}

\author[a,b,c]{Fabiano F. Santos}
\affiliation[a]{Departamento de Física, Universidade Federal do Maranhão, São Luís, 65080-805, Brazil.}
\affiliation[b]{School of Physics, Damghan University, Damghan, 3671641167, Iran.}
\affiliation[c]{Centro de Ciências Exatas, Naturais e Tecnológicas, UEMASUL, 65901-480, Imperatriz, MA, Brazil.}
\emailAdd{fabiano.ffs23@gmail.com}

\abstract{We investigate how the sewing operations familiar from topological quantum field theory provide a useful organizational framework for semiclassical AdS/BCFT path integrals. Our purpose is not to identify AdS/BCFT with a topological field theory: the bulk geometry, the end-of-the-world brane, and the on-shell action remain intrinsically metric-dependent. Rather, we use the TQFT language of state preparation, orientation reversal, and gluing to separate interface data from local geometric contributions. For AdS$_3$/BCFT$_2$, we review how the Euclidean on-shell action and the Ryu--Takayanagi prescription reproduce the Affleck--Ludwig boundary entropy $\log g$ after holographic renormalization. We then formulate the Complexity=Action calculation in the presence of an end-of-the-world brane, carefully distinguishing bulk, brane, joint, and null-boundary contributions. We emphasize that finite terms in a relative Wheeler--DeWitt action depend on the choice of regulator matching and on the null-boundary counterterm scheme unless their scheme independence is established explicitly. Finally, we extend the geometric construction to a planar BTZ background and derive the brane embedding from the Neumann boundary condition. The resulting framework clarifies which aspects of holographic complexity are sensitive to boundary conditions and which require further renormalization input before they can be
interpreted as universal boundary data.}

\begin{document}
	\maketitle
	\newcommand{\limit}[3]
	{\ensuremath{\lim_{#1 \rightarrow #2} #3}}

\section{Introduction} \label{Sec1}

The quest for a deep understanding of the nature of quantum entanglement and gravitational information has been strongly driven by the holographic correspondence \cite{Maldacena:1997re}. Within this framework, the bulk geometry is interpreted as an emergent manifestation of the informational structure of the boundary theory \cite{Witten:1998qj}. The extension of this duality to boundary conformal field theories (BCFTs), known as the AdS/BCFT correspondence \cite{Takayanagi:2011zk, Fujita:2011fp}, provides a particularly well-suited laboratory for investigating systems with interfaces, non-trivial boundary conditions, and localized degrees of freedom. The presence of an end-of-the-world (EOW) brane, denoted by ${\cal Q}$, modifies the physical domain of the geometry and introduces new terms into the gravitational action, thereby affecting observables such as entanglement entropy, pseudo-entropy, and holographic complexity \cite{Ryu:2006bv, Santos:2024cvx, Santos:2025fdp}.

Holographic complexity in the presence of physical boundaries has been investigated previously in several complementary frameworks. In particular, the complexity-fidelity-susceptibility analog of the Affleck--Ludwig g-theorem in AdS$_3$/BCFT$_2$ has been shown to decrease along boundary renormalization-group flows \cite{Flory:2017ftd}. Subsequently, we carried out a systematic analysis of holographic complexity in AdS$_3$/BCFT$_2$ using both the Complexity=Volume and Complexity=Action prescriptions \cite{Braccia:2019xxi}. Their results demonstrate, in particular, that the boundary dependence and UV structure of the complexity are prescription-dependent: while CV-type proposals exhibit boundary-dependent logarithmic divergences, the CA prescription yields a boundary-sensitive finite term. Our purpose is not to rederive these results as new. Instead, we use the gluing language of TQFT as an organizational device for interpreting the localized brane, joint, and counterterm contributions that enter a relative CA observable.

In parallel, the functorial formulation of topological quantum field theory in terms of cobordisms and the Atiyah--Segal axioms provides a precise language for state preparation, amplitude composition, and the gluing of Hilbert spaces  \cite{Simon:2023hdq, Carqueville:2017fmn, Dedushenko:2018aox}. The role of this language in the present work is deliberately limited. Asymptotically AdS gravity is a metric-dependent and dynamical theory, and we do not claim that it defines a TQFT, an extended TQFT, or a functorial representation of the cobordism category. Instead, we use selected TQFT notions as an organizational analogy for the sewing of path integrals. The analogy applies to the combinatorial structure of decompositions, orientations, and contractions of interface data, whereas the weights of the corresponding gravitational configurations remain dynamical and depend on the bulk metric, the brane embedding, the boundary conditions, and the renormalization prescription \cite{Nishioka:2021cxe, Kawamoto:2023wzj}.

In this work, we systematically investigate the intersection between TQFT axioms and the dynamics of AdS/BCFT. Our primary objective is to establish a dictionary between state preparation via holographic propagators and EOW branes on one side, and cobordism, factorization, and contraction operations in the boundary Hilbert space on the other \cite{Melnikov:2025tui, Melnikov:2023wwc, Melnikov:2023nzn}. This dictionary allows for a geometric organization of interface contributions and clarifies how boundary data are incorporated into the composition of gravitational amplitudes.

Our second objective is to analyze holographic complexity within the "Complexity Equals Action" (CA) prescription, emphasizing the separation between universal boundary contributions and local terms dependent on the renormalization scheme \cite{Santos:2025fdp, Brown:2015bva, Lehner:2016vdi, complexityshocks, entrenholo, vidal_tns_geo, Hartman:2013qma, Susskind:2014moa}. In this formulation, complexity is determined by the gravitational action evaluated on the Wheeler--DeWitt (WDW) patch, including volume, boundary, and joint contributions, null counterterms, and, when applicable, brane-localized terms and matter fields. The decomposition of the WDW patch into elementary regions makes the analogy with cobordism composition explicit, while preserving the metric dependence necessary to describe gravitational dynamics.

Beyond the vacuum case, we extend this analysis to thermal geometries and black holes, specifically the planar BTZ black hole with an EOW brane \cite{Takayanagi:2011zk}. The presence of a thermal scale introduces new geometric structures, such as the horizon, the interior region, and potential intersections between the brane and the null boundaries of the WDW patch \cite{Fujita:2011fp}. Consequently, the relevant action receives additional contributions from the brane and the joints located at its intersection with null surfaces. For thermofield double states, the late-time growth of the action reproduces the universal thermal contribution associated with $2M = T_{\rm H} {\cal S}_{\rm BH}$ \cite{Santos:2025fdp, Brown:2015lvg}, while the brane adds a contribution sensitive to the holographic boundary condition. This structure allows for an interpretation of complexity as a quantity that depends not only on the extent of the black hole's thermal interior but also on how the system is terminated at the boundary. By keeping the metric data at the cutoff fixed, the comparison between different boundary conditions isolates an interface contribution that can be identified as the universal component associated with the BCFT.

The same conceptual organization allows for the systematic inclusion of brane-intrinsic matter fields, localized defects of higher codimension, and potential extensions to modified gravity theories \cite{Santos:2025fdp}. In these instances, the composition and gluing structure remains formally analogous to that of a TQFT, but the weights assigned to states and interfaces are determined by an effective action depending on the metric, temperature, matter fields, and coupling parameters. Thus, the topological language does not replace gravitational dynamics; rather, it provides an organizing principle to distinguish composition data, local contributions, and universal boundary information.

The results of this paper can be summarized as follows:
\begin{itemize}
    \item We formulate a structural correspondence between selected TQFT sewing
    operations and the composition of semiclassical AdS/BCFT path integrals.
    This correspondence concerns state preparation, orientation, factorization,
    and interface data; it does not constitute an equivalence between AdS/BCFT
    and a topological quantum field theory.

    \item We show that an end-of-the-world brane is a physical interface whose
    contribution to a gravitational gluing operation is encoded in its induced
    metric, extrinsic curvature, tension, and possible joint and counterterm
    contributions.

    \item We review the extraction of the Affleck--Ludwig boundary entropy from
    the renormalized Euclidean action and from holographic entanglement entropy,
    making explicit the subtraction prescription and the boundary data held fixed.

    \item We formulate the Complexity=Action calculation for Wheeler--DeWitt
    patches intersecting an end-of-the-world brane. We identify the regulator- and
    counterterm-dependent contributions that must be controlled before a finite
    relative complexity can be assigned a universal BCFT interpretation.

    \item We derive the corresponding brane embedding in the planar BTZ geometry
    and specify the assumptions under which brane-localized contributions affect
    the late-time action growth.
\end{itemize}

The structure of this paper is as follows. In Section \ref{Sec2}, we review the Atiyah--Segal axioms for TQFTs and discuss their extension to the AdS/BCFT correspondence, focusing on factorization, interface orientation, and state preparation. In Section \ref{Sec3}, we analyze the excision of worldlines and the decomposition of gravitational geometries into patches, emphasizing the role of joints, boundaries, and counterterms in the Wheeler--DeWitt action used for the complexity calculation. We also discuss how the Affleck--Ludwig entropy, $\log g$, emerges as an interface residue after the separation of local metric contributions \cite{Tang:2017xjc, Harper:2024aku}. In Section \ref{Sec4}, we extend the construction to thermal geometries, presenting the case of the BTZ with an EOW brane and identifying the universal late-time growth contribution alongside the additional interface-associated terms. Finally, in Section \ref{Sec5}, we present our conclusions and discussion.
  
\section{Topological QFT vs AdS/BCFT correspondence}\label{Sec2}

A TQFT in $D+1$ spacetime dimensions is a functor that depends on the topology \cite{Simon:2023hdq,Carqueville:2017fmn,Dedushenko:2018aox}; AdS/BCFT are metric gravitational theories \cite{Shimaji:2018czt,Nishioka:2021cxe,Takayanagi:2011zk,Fujita:2011fp,Santos:2024cwf,Santos:2025ugv,Santos:2025fdp,Santos:2024cvx}. In our description, $\Sigma$ is a $D$-dimensional spatial hypersurface, ${\cal M}$ stands for a $(D+1)$-dimensional cobordism, ${\cal N}$ for the holographic bulk, and ${\cal Q}$ for the End-of-the-World (EOW) brane. In a TQFT, a smooth deformation of a manifold that does not alter its topology leaves ${\cal Z}$ unchanged \cite{Ritz-Zwilling:2023sya}. In gravity, the same deformation can change the area, extrinsic curvature, the position of a joint, or the on-shell action. Accordingly, TQFT identities will be used only as organizing principles for orientation, factorization, and sewing; they will not be interpreted as exact identities for metric-dependent gravitational observables. In this dictionary, orientation reversal, disjoint-union factorization, and the sewing pattern are purely topological operations, whereas the gravitational weights assigned to them—including the on-shell action, brane embedding, extremal areas, joints, and counterterms—depend on the metric and boundary conditions.

The dictionary introduced below has the status of a semiclassical organizational framework. It is neither a rigorous categorical construction nor an assertion that the AdS/BCFT correspondence is an actual functor from a cobordism category to finite-dimensional Hilbert spaces. Its domain of validity is restricted to the use of gluing data as a bookkeeping device for gravitational path integrals in fixed topological sectors and with specified boundary conditions. In particular:

\begin{itemize}
\item the topological side keeps track of orientation, decomposition, factorization, and contraction of interface data;
\item the gravitational side supplies the metric-dependent weights through the Einstein action, brane action, joint terms, null-boundary terms, and counterterms;
\item a change of metric, cutoff prescription, brane tension, or localized matter content can modify the gravitational observable even when the topology remains unchanged.
\end{itemize}
Therefore, the correspondence should be understood as an analogy at the level of sewing structure, supplemented by a semiclassical gravitational variational problem. No topological invariance is attributed to the gravitational action, extremal areas, or Wheeler--DeWitt action.

We shall now present and review some axioms discussed in \cite{Simon:2023hdq,Carqueville:2017fmn,Dedushenko:2018aox}, which will be fundamental to our construction.

\begin{tcolorbox}[colback=red!10, colframe=white!80!black, title={{\color{black}Axiom 1: Hilbert spaces} \cite{Simon:2023hdq,Carqueville:2017fmn,Dedushenko:2018aox}}]
To each closed oriented manifold $\Sigma$, we associate
\begin{equation}
 \Sigma\longmapsto {\cal V}(\Sigma),\qquad \dim {\cal V}(\Sigma)<\infty.
\end{equation}
The dependence is purely topological. For an orthonormal basis $\{\ket{\psi_{\Sigma,\alpha}}\}_{\alpha=1}^{d_\Sigma}$,
\begin{equation} \braket{\psi_{\Sigma,\alpha}|\psi_{\Sigma,\beta}}=\delta_{\alpha\beta},\,\,\mathbb{I}_{V(\Sigma)}=\sum_{\alpha=1}^{d_\Sigma}\ket{\psi_{\Sigma,\alpha}}\bra{\psi_{\Sigma,\alpha}}.
\end{equation}
\end{tcolorbox}

The physical content of this assignment is that spatial topology classifies global low-energy sectors. Non-contractible cycles allow, for example, for fluxes or Wilson lines that cannot be removed by local operations \cite{Shimaji:2018czt}. Consequently, two surfaces with similar local geometry can support state spaces of different dimensions if they have distinct topologies. The sphere $S^2$, without insertions, has no non-contractible cycles; the torus ${\cal T}^2$, on the other hand, possesses two fundamental cycles. In the interpretation of topological phases, the vectors in ${\cal V}(\Sigma)$ describe globally distinguishable sectors, even though no strictly local operator can, in general, connect them.

\begin{tcolorbox}[colback=red!10, colframe=white!80!black, title={{\color{black}Axiom 2: disjoint union and factorization} \cite{Simon:2023hdq,Carqueville:2017fmn,Dedushenko:2018aox}}]
For $\Sigma=\Sigma_1\cup\Sigma_2$,
\begin{equation}\label{tensor}
 {\cal V}(\Sigma_1\cup\Sigma_2)={\cal V}(\Sigma_1)\otimes {\cal V}(\Sigma_2),\qquad {\cal V}(\varnothing)=\mathbb C.
\end{equation}
If $d_i=\dim {\cal V}(\Sigma_i)$, a product basis is $\ket{\alpha, \beta}=\ket{\alpha}_1\otimes\ket{\beta}_2$. Thus,
\begin{equation}
 \dim {\cal V}(\Sigma_1\cup\Sigma_2)=\sum_{\alpha=1}^{d_1}\sum_{\beta=1}^{d_2}1=d_1d_2.
\end{equation}
\begin{center}
\begin{tikzpicture}[>=Latex,scale=1]
\draw[thick] (0,0) ellipse (1.2 and .55); \node at (0,0){$\Sigma_1$};
\draw[thick] (4,0) ellipse (1.2 and .55); \node at (4,0){$\Sigma_2$};
\draw[->,very thick] (1.4,0)--(2.6,0) node[midway,above]{$\cup$};
\node[draw,rounded corners,minimum width=0.6cm,minimum height=.8cm] at (6.6,0) {${\cal V}(\Sigma_1)\otimes {\cal V}(\Sigma_2)$};
\end{tikzpicture}
\end{center}
\end{tcolorbox}

Eq.~\eqref{tensor} expresses locality in its purest form: before introducing any process that connects the two components, the degrees of freedom are independent \cite{Carqueville:2017fmn,Dedushenko:2018aox}. The multiplication of dimensions is also the source of the extensivity of the number of sectors when decoupled topological systems are placed side-by-side. An important physical subtlety is that this factorization applies to the disjoint union. It does not state that a connected spatial region can be freely factored into sub-regions; in gauge theories and gravity, Gauss constraints and boundary data make this latter issue precisely non-trivial.

\begin{tcolorbox}[colback=red!10, colframe=white!80!black, title={{\color{black}Axioms 3--4: states, orientation, and gluing }\cite{Simon:2023hdq,Carqueville:2017fmn,Dedushenko:2018aox}}]
If $\partial{\cal M}=\Sigma$, the topological path integral prepares
\begin{equation}
 {\cal Z}({\cal M})\in {\cal V}(\Sigma).
\end{equation}
For $\partial {\cal M}=\varnothing$, ${\cal Z}({\cal M})\in\mathbb C$. Orientation reversal gives
\begin{equation}\label{orientation}
 {\cal V}(\Sigma^*)={\cal V}(\Sigma)^*.
\end{equation}
\end{tcolorbox}

Writing ${\cal Z}({\cal M})=\sum_\alpha {\cal Z}_\alpha({\cal M})\ket{\alpha}$ and ${\cal Z}({\cal M}')=\sum_\alpha {\cal Z}_\alpha({\cal M}')\ket{\alpha}$, with $\partial {\cal M}=\Sigma$ and $\partial {\cal M}'=\Sigma^*$, the gluing is given by
\begin{align}\label{glue}
 {\cal Z}({\cal M}\cup_\Sigma {\cal M}')&=\bra{{\cal Z}({\cal M}')}{\cal Z}({\cal M})\rangle\\
 &=\sum_{\alpha,\beta}{\cal Z}_\alpha({\cal M}')^*{\cal Z}_\beta({\cal M})\braket{\alpha|\beta}
 =\sum_{\alpha=1}^{d_\Sigma}{\cal Z}_\alpha({\cal M}')^*{\cal Z}_\alpha({\cal M}).
\end{align}
Thus, the geometric identification of the boundary becomes a complete contraction of the interface indices.
\begin{center}
\begin{tikzpicture}[>=Latex,scale=.95]
\draw[fill=blue!10,thick] (0,0).. controls (.4,1.2) and (1.6,1.2)..(2,0)..controls(1.6,-1.2)and(.4,-1.2)..(0,0);
\draw[fill=orange!15,thick] (4,0).. controls (4.4,1.2) and (5.6,1.2)..(6,0)..controls(5.6,-1.2)and(4.4,-1.2)..(4,0);
\draw[very thick] (2,1.05)--(2,-1.05); \draw[very thick] (4,1.05)--(4,-1.05);
\node at (1,0){${\cal M}$}; \node at (5,0){${\cal M}'$}; \node[right] at (2,.7){$\Sigma$}; \node[left] at (4,.7){$\Sigma^*$};
\draw[->,very thick] (2.35,0)--(3.65,0) node[midway,above]{glue};
\node[draw,rounded corners,align=center] at (8,0) {${\cal Z}({\cal M}\cup_\Sigma {\cal M}')$\\$=\bra{{\cal Z}({\cal M}')}{\cal Z}({\cal M})\rangle$};
\end{tikzpicture}
\end{center}

Orientation plays the physical role of distinguishing the state preparation and its conjugate amplitude. A boundary with reversed orientation naturally carries the dual space: in the language of time evolution, a ket prepared in the past becomes a bra when the region is viewed from the opposite side \cite{Melnikov:2025tui,Melnikov:2023wwc,Melnikov:2023nzn,Melnikov:2022qyt}. Eq.~\eqref{glue} is, therefore, the topological version of the composition of path integrals. All degrees of freedom living at the interface $\Sigma$ are summed over; no interface information can remain open after gluing. This observation will be decisive in AdS/BCFT: the brane ${\cal Q}$ cannot be treated merely as a line drawn in the bulk \cite{Takayanagi:2011zk,Fujita:2011fp}, as it specifies variational data and, potentially, degrees of freedom that participate in the sewing \cite{Geng:2022dua}. As we know, according to the works of \cite{Santos:2024cvx,Geng:2022dua,Ryu:2006bv}, the replica trick calculation of entanglement entropy is constructed by gluing $n$ copies of the geometry along a region ${\cal A}$ and its complement. The twist operators $\Phi_n(r,t)$ implement precisely this cyclic identification between replica sheets at the bipartition points. Thus, the two-point correlation function that determines ${\cal S}_{{\cal A}}$ can be interpreted as an amplitude associated with a geometry with interfaces, whose sewing data encode the partition between bulk and boundary channels \cite{Basu:2023jtf}. In this language, the contribution ${\cal S}_{bdry}$ is not merely an additive term in the entropy: it contains information about the degrees of freedom associated with the boundaries or, holographically, with the EOW branes \cite{Fujita:2011fp}.

The observation above motivates the use of TQFT language. We do not intend to identify the metric gravitational theory of AdS/BCFT with a strict TQFT, as quantities such as extremal surface areas, extrinsic curvatures, and on-shell actions are metric-dependent. However, TQFT axioms provide a precise organizing framework to distinguish purely topological sewing data—state spaces \cite{Simon:2023hdq}, orientation, factorization, and interface contraction—from the geometric and dynamical data that characterize the holographic realization \cite{Melnikov:2022qyt}. In particular, gluing along a common hypersurface will be the abstract analog of the identification between bulk regions, BCFT boundaries, and EOW branes that appears in the replica entropy calculation. In this way, we can summarize: Cobordisms, Gluing, and the Replica Trick as follows:

\begin{tcolorbox}[colback=red!10, colframe=white!80!black, title={{\color{black}Cylinder $\rightarrow$ Torus: Identity and Trace} \cite{Simon:2023hdq,Melnikov:2022qyt,Melnikov:2023wwc}}]
\[
\Sigma \times \mathbb I
\]
\[
{\cal Z} = \mathbb I_{{\cal V}_\Sigma} = \sum_\alpha |\alpha\rangle\langle\alpha|
\]

\begin{itemize}
    \item $|\alpha\rangle$: outgoing state
    \item $\langle\alpha|_{\Sigma^*}$: incoming state
\end{itemize}

$\downarrow$ Identify:
\[
\Sigma \cong \Sigma^*
\]

\subsection*{Torus}

\[
\Sigma \times S^1
\]
\[
{\cal Z} = \operatorname{Tr} \mathbb I = \dim {\cal V}(\Sigma)
\]
\end{tcolorbox}

Trivial evolution preserves every sector. Identifying $\Sigma \cong \Sigma^*$ closes the interval, converting a sum into a trace. The reversed orientation of $\Sigma^*$ distinguishes $\langle\alpha|$ from $|\alpha\rangle$: preparation versus conjugate amplitude.

\begin{tcolorbox}[colback=red!10, colframe=white!80!black, title={{\color{black}Gluing Cobordisms: Composition of Path Integrals} \cite{Simon:2023hdq,Melnikov:2022qyt,Melnikov:2023wwc}}]
\[
{\cal Z}({\cal M}_1): \Sigma_1 \rightarrow \Sigma
\]
\[
{\cal Z}({\cal M}_2): \Sigma \rightarrow \Sigma_2
\]

Gluing along the shared boundary $\Sigma$ gives:
\[
{\cal Z}({\cal M}_1 \cup_\Sigma {\cal M}_2) = {\cal Z}({\cal M}_2) \cdot {\cal Z}({\cal M}_1)
\]

All degrees of freedom on $\Sigma$ are integrated out---no interface information remains open.
\end{tcolorbox}

\begin{tcolorbox}[colback=red!10, colframe=white!80!black, title={{\color{black}Replica Trick: Gluing $n$ Copies and Entanglement Entropy} \cite{Simon:2023hdq,Melnikov:2022qyt,Melnikov:2023wwc}}]
\begin{itemize}
    \item Sheet 3 $\bullet\ \Phi_n$
    \item Sheet 2 $\bullet\ \Phi_n$
    \item Sheet 1 $\bullet\ \Phi_n$
\end{itemize}

\[
{\cal Z}({\cal M}_n) = \operatorname{Tr} \rho_{A}^n
\]
\end{tcolorbox}

The $n$ copies are glued cyclically along:
\[
{\cal A} \subset \Sigma
\]

After analytic continuation $n \rightarrow 1$:
\[
{\cal S}_{{\cal A}} = -\left.\partial_n \log Z({\cal M}_n)\right|_{n=1}
\]

With an RT surface and an end-of-the-world (EOW) brane \cite{Basu:2023jtf}:
\[
{\cal S}_{\cal A} = {\cal S}_{\mathrm{bulk}} + {\cal S}_{\mathrm{bdry}}
\]
Here, ${\cal S}_{\mathrm{bdry}}$ encodes degrees of freedom associated with the EOW brane ${\cal Q}$.

\begin{tcolorbox}[colback=red!10, colframe=white!80!black, title={{\color{black}Conceptual Flow} \cite{Simon:2023hdq,Melnikov:2022qyt,Melnikov:2023wwc}}]
\[
\Sigma \times \mathbb I
\quad\longrightarrow\quad
\Sigma \times S^1
\quad\longrightarrow\quad
{\cal M}_n\ \text{(replicas)}
\quad\longrightarrow\quad
{\cal S}_{{\cal A}}
\]
More explicitly:

\begin{enumerate}
    \item \textbf{Cylinder}
    \[
        {\cal Z} = \mathbb I
    \]

    \item \textbf{Close the cylinder $\rightarrow$ Torus}
    \[
        {\cal Z} = \operatorname{Tr} \mathbb I = \dim {\cal V}
    \]

    \item \textbf{Take $n$ copies $\rightarrow$ Replica manifold}
    \[
        {\cal Z} = \operatorname{Tr} \rho^n, \qquad \text{twist } \Phi_n
    \]

    \item \textbf{Continue $n \rightarrow 1$ $\rightarrow$ Entropy with EOW brane}
    \[
        {\cal S}_{{\cal A}} = {\cal S}_{\mathrm{bulk}} + {\cal S}_{\mathrm{bdry}}
    \]
\end{enumerate}
\end{tcolorbox}

Topological invariance guarantees that $\dim {\cal V}(\Sigma)$ is an invariant. The replica trick inherits this sewing structure, with brane ${\cal Q}$ encoding nontrivial variational data at the interface.

The cobordism $\Sigma \times \mathbb{I}$ has an incoming and an outgoing boundary; therefore,
\begin{equation}
 {\cal Z}(\Sigma \times \mathbb{I})=\sum_\alpha\ket{\alpha}\bra{\alpha}=\mathbb{I}_{V(\Sigma)}.
\end{equation}
By gluing the ends of the cylinder, one obtains $\Sigma \times S^1$. From the definition of the trace,
\begin{equation}\label{trace}
 {\cal Z}(\Sigma \times S^1)=Tr\mathbb{I}=\sum_\alpha\braket{\alpha|\alpha}=d_\Sigma=\dim {\cal V}(\Sigma).
\end{equation}
In particular, in the example discussed by \cite{Simon:2023hdq}, $\dim {\cal V}(S^2)=1$, while $\dim {\cal V}(T^2)$ counts the species of particles/anyons.
\begin{center}
\begin{tikzpicture}[>=Latex]
\draw[thick] (0,0) ellipse (1 and .35); \draw[thick] (0,2.2) ellipse (1 and .35);
\draw[thick] (-1,0)--(-1,2.2); \draw[thick] (1,0)--(1,2.2);
\node at (0,1.1) {$\Sigma\times\mathbb I$}; \node[right] at (1,2.2){$\Sigma$}; \node[right] at (1,0){$\Sigma^*$};
\draw[->,very thick] (2,1.1)--(3.5,1.1) node[midway,above]{identify};
\draw[thick] (5,1.1) ellipse (1 and 1.25); \draw[dashed] (4,1.1) arc (180:360:1 and .3);
\node at (5,1.1) {$\Sigma\times S^1$};
\node at (7.5,1.1) {${\cal Z}=Tr\mathbb I=\dim {\cal V}(\Sigma)$};
\end{tikzpicture}
\end{center}

The cylinder represents evolution without non-trivial topological dynamics: it exactly preserves each sector of the Hilbert space \cite{Simon:2023hdq,Melnikov:2022qyt,Melnikov:2023wwc}. By closing the direction of the interval, the sum over states compatible with the identification transforms the amplitude into a trace. Eq.~\eqref{trace} demonstrates a striking physical characteristic of TQFTs: a partition function on a compact manifold can directly measure the ground-state degeneracy \cite{Melnikov:2022qyt,Melnikov:2023wwc,Basu:2023jtf}. This contrasts with an ordinary local quantum field theory, where the thermal trace depends on the length scale of the Euclidean circumference and the energy of excitations. Here, the absence of such a scale is precisely the signature of topological invariance.

\section{Worldlines, excision, and solid torus}\label{Sec3}

With the perspective of the Topological QFT vs. AdS/BCFT correspondence, we can perform the transition from TQFT axioms to the decomposition of gravitational geometries, which can be stated as follows: the excision of a tubular region and the gluing of cobordisms show that data initially interpreted as bulk insertions can be reorganized as states or boundary conditions on an interface \cite{Simon:2023hdq,Melnikov:2022qyt,Melnikov:2023wwc}. In AdS/BCFT \cite{Takayanagi:2011zk,Fujita:2011fp}, this interface is geometrically realized by the EOW brane $\cal Q$, whose presence not only modifies the effective topology of the regions to be glued but also introduces additional metric and variational data \cite{Santos:2024cwf}. Thus, while in a TQFT the sewing is expressed by an exact contraction in the Hilbert space associated with the common boundary \cite{Shimaji:2018czt,Nishioka:2021cxe}, in gravity the same operation must be accompanied by the analysis of the action terms localized at the interface — including Gibbons--Hawking--York contributions, joints, counterterms, and the action of the brane itself \cite{Chen:2020uac,Chen:2020hmv,Geng:2021mic,Geng:2021hlu}. The following discussion formalizes this distinction by comparing the topological excision of worldlines with the decomposition of gravitational patches, identifying precisely which contributions cancel out for artificial interfaces and which survive as physical data associated with the sewing. As discussed by \cite{Simon:2023hdq,Melnikov:2022qyt,Melnikov:2023wwc}, a worldline labeled $a$ has a tubular neighborhood $S^1 \times D^2$. After removing it, the new boundary is ${\cal T}^2$ and the label is represented by a vector $\ket{L_a} \in {\cal V}({\cal T}^2)$. If $X = {\cal M} \setminus (S^1 \times D^2)$, then the amplitude is the gluing
\begin{equation}
 {\cal Z}({\cal M};a)=\bra{{\cal Z}(X)}L_a\rangle.
\end{equation}
This operationally demonstrates the transition "bulk insertion $\leftrightarrow$ boundary data" \cite{Simon:2023hdq}. By gluing two solid tori with the modular transformation that swaps meridian and longitude,
\begin{equation}\label{Smat}
 {\cal S}_{ab}={\cal Z}\big(S^3;\text{Hopf's link labeled by}a,b\big).
\end{equation}

\begin{center}
\begin{tikzpicture}[>=Latex,scale=.95]
\draw[thick,fill=gray!10] (0,0) ellipse (1.4 and .55); \draw[thick] (-1.4,0)--(-1.4,2); \draw[thick] (1.4,0)--(1.4,2); \draw[thick,fill=gray!10] (0,2) ellipse (1.4 and .55);
\draw[red,very thick] (0,1.05) ellipse (.48 and .7); \node[red] at (.8,1.05){$a$}; \node at (0,-.8){$S^1\!\times\!D^2$};
\draw[->,very thick] (2.3,1)--(4,1) node[midway,above]{excise};
\draw[thick] (5.4,1) ellipse (1.25 and .62); 
\draw[dashed] (4.15,1) arc (180:360:1.25 and .62); \node at (5.4,1){$\partial X={\cal T}^2$};
\node[draw,rounded corners,align=center] at (8.4,1) {$\ket{L_a}\in {\cal V}({\cal T}^2)$};
\end{tikzpicture}
\end{center}

The excision makes visible a principle of "topological holography" in a narrow sense \cite{Moradi:2022lqp,Wang:2018edf}: the type of a point-like excitation in the bulk of a slice can be encoded by an admissible state on the boundary surrounding its worldline. For phases with anyons, the labels $a,b$ describe superselection sectors, and the matrix ${\cal S}_{ab}$ from Eq.~\eqref{Smat} measures the mutual braiding of their trajectories. This construction does not claim that all information in a gravitational theory is boundary-based; it provides a precise analogy for understanding why defects, Wilson lines, and boundary conditions are essential data when gluing regions \cite{Arcioni:2002vv}.

The analogy established by excision suggests a concrete gravitational formulation for the sewing operation. In a TQFT \cite{Wang:2018edf}, removing a neighborhood of a worldline transforms the bulk defect into boundary data: the state $\ket{L_a} \in \mathcal{V}(T^2)$ with ${\cal Z}({\cal M};a)=\bra{{\cal Z}(X)}L_a\rangle$ contains precisely the information that must be specified so that the excised region can be subsequently re-glued \cite{Nishioka:2021cxe}. In this formalism, gluing is a complete contraction over the interface degrees of freedom. In gravity, the decomposition of a region $\mathcal{W}$ into patches $\mathcal{W}_L$ and $\mathcal{W}_R$ realizes the same structural idea, but with an essential difference: the interface possesses its own metric and variational data \cite{Nishioka:2021cxe,Takayanagi:2011zk,Fujita:2011fp,Geng:2022dua,Ryu:2006bv}.

If a smooth internal surface $\Gamma$ is introduced merely as a decomposition device, these data should not produce additional physical content, and the cancellation between the Gibbons--Hawking--York (GHY) contributions from both sides ensures that the action of the recomposed geometry is simply additive \cite{Dordevic:2024ziw}. On the other hand, when an EOW brane $\cal Q$ is included, a joint, a signature transition, or an effective change in boundary conditions, it ceases to be an auxiliary boundary: much like the torus surrounding a worldline in TQFT, the interface begins to carry physical information that must be included in the gluing operation. Suppose that $\mathcal W=\mathcal W_L\cup_\Gamma\mathcal W_R$ can be separated into volume integrals:
\begin{equation}
 I_{\rm bulk}[\mathcal W]=I_{\rm bulk}[\mathcal W_L]+I_{\rm bulk}[\mathcal W_R].
\end{equation}
The GHY contributions \cite{Dordevic:2024ziw} from a smooth internal interface cancel out because the normals are opposite and $K_R=-K_L$:
\begin{equation}
 I_{\Gamma,L}+I_{\Gamma,R}=\frac1{8\pi G}\int_\Gamma\sqrt{|h|}\,(K_L+K_R)=0.
\end{equation}
When there is a change in boundary type or a non-smooth junction, what remains is
\begin{equation}\label{defectglue}
 \Delta_\Gamma I\equiv I[\mathcal W]-I[\mathcal W_L]-I[\mathcal W_R]
 =I_{\rm joint}(\Gamma)+I_{\rm ct}(\Gamma)+I_{\cal Q}(\Gamma),
\end{equation}
with terms present only when applicable. Eq. \eqref{defectglue} is the gravitational version of the gluing idea \cite{Kawamoto:2023wzj}: it is not an inner product, but an additive law corrected by localized data. The terms $I_{\text{joint}}(\Gamma)$, $I_{\text{ct}}(\Gamma)$, and $I_{\cal Q}(\Gamma)$ in Eq. \eqref{defectglue} thus constitute the gravitational realization of the localized data at the interface; they replace the purely topological contraction of states with a composition rule that depends on the metric and local geometry.
\begin{center}
\begin{tikzpicture}[>=Latex,scale=.95]
\fill[blue!10] (0,0)--(2.8,1.5)--(0,3)--cycle; \fill[orange!15] (2.8,1.5)--(5.6,0)--(5.6,3)--cycle;
\draw[thick] (0,0)--(2.8,1.5)--(0,3)--cycle; \draw[thick] (2.8,1.5)--(5.6,0)--(5.6,3)--cycle;
\draw[very thick,dashed] (2.8,1.5)--(5.6,3); \node at (1.1,1.5){$\mathcal W_L$}; \node at (4.5,1.5){$\mathcal W_R$}; \node[above] at (4.25,2.35){$\Gamma$};
\node[draw,rounded corners,align=center] at (8.5,1.5) {$I[\mathcal W]=I[\mathcal W_L]+I[\mathcal W_R]$\\$+I_{\rm joint}+I_{\rm ct}+I_{\cal Q}$};
\end{tikzpicture}\label{pathw}
\end{center}

When $\Gamma$ is a smooth internal interface, the GHY cancellation expresses that the decomposition has introduced an artificial boundary that should not carry its own physics \cite{Erdmenger:2023hne}. When the interface coincides with a brane, a null joint, or a signature change, the situation changes: there is genuine local information associated with the way the pieces meet. Eq.~\eqref{defectglue} makes this fact explicit. It is the gravitational analogue—richer and metric—of the contraction of indices in \eqref{glue}. A concrete physical objective is to classify which parts of $\Delta_\Gamma I$ are universal under regulator changes and which depend on choices of normalization, tension, or matter couplings. In this sense, in the $AdS/BCFT$ construction \cite{Takayanagi:2011zk,Fujita:2011fp}, the bulk boundary is
\begin{equation}\label{boundaryN}
 \partial {\cal N}={\cal M}\cup {\cal Q},\,\, \partial {\cal M}=\partial {\cal Q}={\cal P}.
\end{equation}
A minimal Euclidean action, omitting counterterms in ${\cal M}$, is
\begin{equation}\label{adsbcftaction}
 I_E=-\frac1{16\pi G}\int_{{\cal N}}\sqrt g\,(R-2\Lambda)
 -\frac1{8\pi G}\int_{{\cal Q}}\sqrt h\,(K-T).
\end{equation}
By varying the induced metric $h_{ab}$ in ${\cal Q}$ (and using the Dirichlet condition in ${\cal M}$),
\begin{equation}
 \delta I_E\big|_{{\cal Q}}=-\frac1{16\pi G}\int_{{\cal Q}}\sqrt h\,[K_{ab}-(K-T)h_{ab}]\,\delta h^{ab}.
\end{equation}
Stationarity with respect to an arbitrary $\delta h^{ab}$ implies
\begin{equation}\label{neumann}
 K_{ab}-(K-T)h_{ab}=0.
\end{equation}
Eq.~\eqref{boundaryN} geometrically realizes the fact that a $BCFT$ has a spatial boundary. The surface ${\cal Q}$ is not an additional asymptotic boundary where a source is freely fixed; it is a dynamic boundary of the bulk, whose position is determined by a Neumann-type condition \cite{Takayanagi:2011zk}. Eq.~\eqref{neumann} is a local balance between the extrinsic curvature of the brane and its tension. Physically, $T$ controls how much the brane "tilts" or carves out the $AdS$ space. This metric dependence is exactly the missing ingredient in a pure $TQFT$ and explains why the dictionary presented in \cite{Simon:2023hdq,Melnikov:2022qyt,Melnikov:2023wwc} should be used as a framework for sewing \cite{Kawamoto:2023wzj}, rather than as a literal equivalence of theories. For example, consider
\begin{equation}
 \dd s^2=\frac{L^2}{z^2}(\dd z^2+\dd x^2+\dd\tau^2).\label{ads3metric}
\end{equation}
The unit normal can be chosen as $n_\mu=(L/z)(-\cos\theta,\sin\theta,0)$ in the basis $(z,x,\tau)$, up to a global sign. The induced metric is
\begin{equation}
 ds_{\cal Q}^2=\frac{L^2}{z^2}(\csc^2\theta\,dz^2+d\tau^2).
\end{equation}
The calculation of $K_{ab}=h_a{}^\mu h_b{}^\nu\nabla_\mu n_\nu$ gives a completely umbilical surface,
\begin{equation}
 K_{ab}=\frac{\cos\theta}{L}h_{ab},\qquad K=\frac{2\cos\theta}{L},
\end{equation}
with adjustable sign depending on the orientation of $n$. Substituting in \eqref{neumann},
\begin{equation}
 \frac{\cos\theta}{L}h_{ab}-\left(\frac{2\cos\theta}{L}-T\right)h_{ab}=0
 \quad\Longrightarrow\quad T=\frac{\cos\theta}{L}\footnote{Here $T$ is the brane tension where we define $T\equiv\frac{{\cal T}_{phys}}{8\pi\, G}=\frac{\tau}{8\pi\,L\,G}=\frac{\sin(\theta_0)}{8\pi\,L\,G}$ with $\tau\equiv\,L{\cal T}_{phys}$, ${\cal T}_{phys}=\frac{\cos(\theta)}{L}$, $\tau=\sin(\theta_0)$, and $\theta_0=\pi/2-\theta$.}.
\end{equation}
Thus, tension is a continuous geometric parameter; it is not, in itself, a topological invariant.
\begin{center}
\begin{tikzpicture}[>=Latex,scale=1]
\fill[blue!8] (0,0)--(6,0)--(6,3.5)--(0,3.5)--cycle;
\draw[very thick] (0,0)--(6,0) node[right]{${\cal M}$};
\draw[red,very thick] (1.1,0)--(4.4,3.5) node[above]{${\cal Q}:\ x=z\cot\theta$};
\draw[->] (1.1,0)--(1.9,0) node[midway,below]{$x$}; \draw[->] (1.1,0)--(1.1,1) node[left]{$z$};
\node at (5.1,2.6) {${\cal N}\simeq\,AdS_3$}; \node[red] at (1.5,.55) {$\theta$};
\end{tikzpicture}
\label{BRAN}
\end{center}

The condition of total umbilicity means that the extrinsic curvature is proportional to the induced metric: \underline{there is no intrinsically preferred direction on} ${\cal Q}$ \cite{Takayanagi:2011zk,Fujita:2011fp}. In the dual of $BCFT_2$, the choice of $\theta$ selects a boundary condition and modifies universal boundary observables, such as the Affleck--Ludwig entropy \cite{Tang:2017xjc}. Even in this highly symmetric example, the physics is no longer topological: varying $\theta$ changes the bulk geometry and, in Lorentzian problems, can alter the intersections of the brane with extremal surfaces or with the $WDW$ patch. The usefulness of this case lies in providing \underline{a simple family of boundary conditions} with which renormalized differences can be defined \underline{without changing the global topology}.

\subsection{Collage Dictionary: Preparation, Contraction, and Holographic Dynamics}
The extension of the axioms proposed by \cite{Simon:2023hdq} is labeled by boundary conditions:
\begin{equation}
 {\cal V}(\Sigma)\ \leadsto\ {\cal V}(\Sigma;\mathcal B),\qquad \mathcal B\leftrightarrow {\cal Q}.
\end{equation}
An integral Euclidean path with condition $\mathcal B$ prepares $\ket{B_{\mathcal B}}$. The collage that closes the geometry produces
\begin{equation}
 {\cal Z}_{\rm disk}=\braket{0|B_{\mathcal B}}\equiv g_{\mathcal B},\qquad {\cal S}_{\rm bdry}=\log g_{\mathcal B}.
\end{equation}
This is a structural analogy with \eqref{glue}; the dependence of \(g\) on \(T\) requires holographic dynamics and does not follow from the topological axioms \cite{Harper:2024aku}. The quantity \(\mathcal B\) emphasizes that a boundary condition is additional physical data, rather than merely the location of a surface. It can specify how fields reflect, which symmetries survive at the boundary, and which defects are allowed to terminate there \cite{Santos:2024cwf,Santos:2025ugv,Geng:2022dua,Ryu:2006bv,Chen:2020uac,Chen:2020hmv}. The state \(\ket{B_{\mathcal B}}\) encodes these properties in the closed-channel description of the \(CFT\). Its overlap with the vacuum, \(g_{\mathcal B}\), measures the universal contribution of the boundary to the entropy \cite{Harper:2024aku,Tang:2017xjc}. The TQFT construction provides an abstract model for the composition of states and amplitudes associated with an interface. In the present setting, however, this comparison is structural rather than literal. The holographic path integral depends on the metric, on the brane embedding, and on the boundary terms required by the variational principle. Consequently, the TQFT language organizes the sewing data, whereas the dependence of $g_{\mathcal B}$ on the boundary condition must be obtained from the dynamical AdS/BCFT calculation. Thus, in order to evaluate this state preparation and contraction, we compute \(g\) by evaluating the partition function, \(Z=e^{-I_E}\), for a \(BCFT\) \cite{Takayanagi:2011zk,Fujita:2011fp,Santos:2024cwf,Santos:2025ugv}, on a disk of radius \(r_D\), as follows:

\begin{equation}
 I_E(\rho_\ast)=\frac{R}{4G_N}\left[
 \frac{r_D^2}{2\epsilon^2}
 +\frac{r_D\sinh(\rho_\ast/R)}{\epsilon}
 +\log\!\left(\frac{\epsilon}{r_D}\right)-\frac{1}{2}
 -\frac{\rho_\ast}{R}
 \right].
 \label{eq:disk-action}
\end{equation}
The terms that depend on the cutoff \(\epsilon\) are removed by local counterterms. To extract the boundary contribution unambiguously, one subtracts the reference configuration at \(T=0\), for which \(\rho_\ast=0\). Thus,
\begin{equation}
 I_E(\rho_\ast)-I_E(0)=-\frac{\rho_\ast}{4G_N}.
 \label{eq:action-difference}
\end{equation}
Since $Z=e^{-I_E}$ and $g_{\mathcal B}=\langle 0 \vert B_{\mathcal B} \rangle$, it follows that
\begin{equation}
 \boxed{\;
 \log g_{\mathcal B}={\cal S}_{bdry}=\frac{\rho_\ast}{4G_N}=\frac{R}{4G_N}arc\tanh(RT)\; }.
 \label{eq:boundary-entropy}
\end{equation}
Using the Brown–Henneaux relation \cite{Brown:1986nw,Terashima:2000gb}, $c=3R/(2G_N)$, we obtain the following intrinsically two-dimensional form:
\begin{equation}
 \boxed{\;
 \log g_{\mathcal B}=\frac{c}{6}\operatorname{arctanh}(RT)\; }.
 \label{eq:g-central-charge}
\end{equation}
In particular, in the limit $|RT|\ll1$,
\begin{equation}
 \log g_{\mathcal B}=\frac{c}{6}\left[RT+\frac{(RT)^3}{3}+\frac{(RT)^5}{5}+\cdots\right].
\end{equation}
A consistency check can be performed via the entanglement entropy, considering a $BCFT$ defined on the half-line and an interval ${\cal A}$ of length $l$ terminating at the boundary. The holographic Ryu--Takayanagi formula \cite{Ryu:2006bv} is given by
\begin{equation}
 {\cal S}_{\cal A}=\frac{{\cal A}(\gamma_{\cal A})}{4G_N}.
 \label{eq:RT}
\end{equation}
The relevant geodesic traverses $-\infty < \rho \leq \rho_\ast$. Its divergent part reproduces the standard $CFT$ contribution, while the endpoint at ${\cal Q}$ contributes the finite term
\begin{equation}
 {\cal S}_{{\cal A}}=\frac{c}{6}\log\!\left(\frac{l}{\epsilon}\right)+\frac{\rho_\ast}{4G_N},
\end{equation}
or, equivalently,
\begin{equation}
 \boxed{\;
 {\cal S}_{{\cal A}}=\frac{c}{6}\log\!\left(\frac{l}{\epsilon}\right)+\log g_{\mathcal B}\; }.
 \label{eq:ee-g}
\end{equation}
Additive constants within the logarithm depend on the choice of regulator convention; the universal term $\log g_{\mathcal B}$ does not \cite{Harper:2024aku}. A comparison of \eqref{eq:boundary-entropy} with \eqref{eq:ee-g} shows that the amplitude of the boundary state coincides with the finite contribution of the geodesic anchored at ${\cal Q}$. This constitutes a structural analogy with \eqref{glue} \cite{Kawamoto:2023wzj}: a geometry with a boundary prepares a vector, while a second geometry allows for its contraction. In the $CFT$, the boundary condition $\mathcal B$ is encoded in the closed-channel state $\ket{B_{\mathcal B}}$, and its overlap with the vacuum defines
\begin{equation}
 g_{\mathcal B}=\langle0\vert B_{\mathcal B}\rangle,\,\, {\cal S}_{bdry}=\log g_{\mathcal B}.
\end{equation}
The \(TQFT\) interpretation captures precisely this structure of states and contractions \cite{Carqueville:2017fmn}. However, it does not, by itself, determine how the weight \(g_{\mathcal B}\) depends on the microscopic parameters of the boundary condition \cite{Dedushenko:2018aox}.

The holographic description supplies this dynamical ingredient. Rather than identifying \(\mathcal B\) solely with the location of a surface, it associates \(\mathcal B\) with physical data on \({\cal Q}\): tension, localized fields, reflection conditions, preserved symmetries, and rules governing the termination of defects \cite{Harper:2024aku,Santos:2024cwf,Santos:2025ugv,Geng:2022dua,Ryu:2006bv,Chen:2020uac,Chen:2020hmv}. The conceptual chain is as follows:
\begin{equation}
 \mathcal B\ \longrightarrow\ {\cal Q}_{\mathcal B}\ \longrightarrow\ \rho_\ast(T_{\mathcal B})
 \ \longrightarrow\ I_E^{\mathrm{ren}}\ \longrightarrow\ \log g_{\mathcal B}.
\end{equation}
Therefore, the dependence of \(g\) on \(T\) requires solving the gravitational equations and evaluating the on-shell action; it does not follow solely from the topological axioms \cite{Harper:2024aku}. The holographic formulation turns the abstract statement that a boundary prepares a state into a concrete computation: the state preparation carries a weight determined by the dynamical geometry of \({\cal Q}\).
\subsection{WDW Patch Action and Joint Prescription}

The $Complexity=Action$ prescription requires the complete gravitational action on the Wheeler--DeWitt patch \cite{Carqueville:2017fmn,Dedushenko:2018aox,Shimaji:2018czt}. In the presence of an end-of-the-world brane, this action contains bulk, brane, joint, and null-boundary contributions. A finite comparison between configurations with different brane tensions is meaningful only after the regulating surfaces have been matched by fixing the same induced boundary data. Holding a coordinate cutoff fixed is generally insufficient, because
the brane embedding changes the relation between the coordinate cutoff and the proper cutoff on the regulated boundary. We therefore distinguish two separate issues. First, the standard null counterterm is required to remove the dependence of the action on the parametrization of the
null generators. Second, finite terms that remain after a relative subtraction must be tested for independence of the matching prescription and of the counterterm
scale $\ell_{\rm ct}$. Only terms that pass both tests may be candidates for a universal boundary contribution. In this sense, for a boundary test, we can evaluate the late-time growth, which for stationary solutions is linear \cite{Susskind:2014rva,Brown:2015bva,Lloyd:2000cry,Brown:2015lvg,Susskind:2018fmx,Brown:2018bms,Brown:2017jil,Brown:2019whu,Brown:2022rwi}:
\begin{equation}\label{complex}
 \frac{d{\cal C}}{dt}=\frac{1}{\pi\hbar}\frac{d I_{WDW}}{dt}=\frac{1}{\pi\hbar}\left(\frac{dI_{\rm bulk}}{dt}+\frac{dI_{\cal Q}}{dt}+\frac{dI_{\rm ct}}{dt}+\frac{dI_{\rm joint}}{dt}+\frac{dI_{\rm null}}{dt}\right).
\end{equation}
For a certain class of solutions, this growth is proportional to ${\cal ST}$ \cite{Santos:2025fdp}, provided the effective characteristic cone \footnote{A detailed description of the characteristic cone was presented for Horndeski gravity in \cite{Santos:2025fdp}} is compatible with the null structure used in the Wheeler-DeWitt ($WDW$) patch. The \emph{Complexity equals Action} conjecture ($CA$-\ref{complex}) proposed by Brown, Roberts, Susskind, Swingle, and Zhao~\cite{Brown:2015bva,Brown:2015lvg} posits that the quantum complexity of the boundary state is proportional to the action evaluated on the $WDW$ patch.

In spaces with a boundary ($BCFT$), the holographic dual introduced by Takayanagi \cite{Takayanagi:2011zk} extends the standard $AdS/CFT$ framework through an EOW brane ${\cal Q}$ that truncates the $AdS$ space and whose tension $T$ controls the boundary entropy $\log g$. A natural question arises: how does the presence of ${\cal Q}$ affect holographic complexity? A complete answer requires the evaluation of the $WDW$ action. We shall work with the vacuum geometry to expose the essential structure. Following \cite{Takayanagi:2011zk}, we introduce a hypersurface ${\cal Q}\subset AdS_3$, the EOW brane, whose equation is given by
\begin{equation}\label{braneEq}
  x = -z\tan\theta_0,\quad \theta_0\in\left(0,\tfrac\pi2\right),
\end{equation}
or equivalently \(x/z = -\tan\theta_0\), which defines a plane passing through the conformal boundary \((\partial AdS_3: z=0)\) at \(x=0\) and tilting into the interior. The brane partitions \(AdS_3\) into the physical region
\begin{equation}
  \mathcal{M}:\; x > -z\tan\theta_0,
\end{equation}
such that the conformal boundary consists of two parts, $\partial\mathcal{M} = \partial AdS_3^{+}\cup {\cal Q}$, where $\partial AdS_3^{+} = \{z=0,\, x>0\}$ and ${\cal Q}$ is the brane. The outward-pointing unit normal (with respect to the region $\mathcal{M}$) to the brane defined by $F(x,z)\equiv x + z\tan\theta_0 = 0$ is
\begin{equation}\label{normalQ}
  n_\mu = \frac{\partial_\mu F}{\sqrt{g^{\mu\nu}\partial_\mu F\,\partial_\nu F}}
  = \frac{z}{L}\left(\cos\theta_0,\,0,\,\sin\theta_0\right),
\end{equation}
with components in the basis \((t,x,z)\). The extrinsic curvature of \({\cal Q}\) is
\begin{equation}\label{Kbrane}
  K_{ij} = -\frac12 L_n g_{ij}\big|_{\cal Q},\quad
  K \equiv g^{ij}K_{ij},
\end{equation}
which yields
\begin{equation}\label{KQ}
  K\big|_{\cal Q} = \frac{2\sin\theta_0}{L}.
\end{equation}
The Neumann condition that determines the equilibrium position of the brane is given by
\begin{equation}\label{Neumann}
  K_{ij} - K\,h_{ij} = 8\pi G\, T_{\rm brane}\,h_{ij},
\end{equation}
where \(T_{\rm brane}\) denotes the intrinsic tension and \(h_{ij}\) is the induced metric. For the static brane described by \eqref{braneEq}, this reduces to
\begin{equation}\label{TvsTheta}
  T_{\rm brane} = \frac{\sin\theta_0}{8\pi G\, L} \equiv \frac{T}{8\pi G\,L},
\end{equation}
where we have defined \(T \equiv \sin\theta_0\) to simplify the notation. For a Lorentzian region \(\mathcal{W}\) (the WDW patch), the total action is expressed as

\begin{eqnarray}\label{IW}
&&I_{WDW} = I_{bulk} + I_{\rm nonnull} + I_{\rm null} + I_{\rm joint}+ I_{\rm ct} + I_{\cal Q}.\\
&&I_{bulk} = \frac{1}{16\pi G}\int_{\mathcal{W}}\!\sqrt{-g}\,d^3x\, \bigl(R - 2\Lambda\bigr),\\
&& I_{\rm nonnull} = \frac{\epsilon}{8\pi G}\int_{\partial\mathcal{W}\setminus\mathcal{N}}\!\sqrt{|h|}\,d^2x\,K,\,\, \epsilon=+1\,, space\,, and\,-1\,,time,\\
&& I_{\rm joint} = \frac{1}{8\pi G}\int_{\mathcal{J}}\!\sqrt{\sigma}\,a\,d\theta,\,\,a = \log\!\left|\frac{k\cdot\bar k}{2}\right|,\\
&&  I_{\rm ct} = \frac{1}{8\pi G}\int_{\mathcal{N}}\!\sqrt{\gamma}\,\Theta\,\log(lct\,\Theta)\,d\lambda\,d\theta,\,\,\Theta = \partial_\lambda\log\sqrt{\gamma},\\
&&  I_{\cal Q} = \frac{1}{8\pi G}\int_{{\cal Q}}\!\sqrt{|h_{\cal Q}|}\,d^2x\,(K_{\cal Q} - T_{\rm brane}),
\end{eqnarray}
The bulk integral measures the dynamics within the causally relevant interior; the boundary terms render the variational principle well-posed; the joints account for the non-smooth intersections of boundary components; and the null counterterms eliminate ambiguities in the normalization of the generators \cite{Dedushenko:2018aox,Shimaji:2018czt}. In particular, a \(CA\) analysis that retains only \(I_{bulk}\) is not invariant under the choices required to define the patch \cite{Susskind:2014rva,Brown:2015bva,Lloyd:2000cry,Brown:2015lvg,Susskind:2018fmx}. The presence of the EOW brane introduces \(I_{\cal Q}\) and creates new intersections between \({\cal Q}\) and the null sheets. We now proceed to compute the contributions to \(I_{WDW}\), starting with the action  $I=I_{bulk}+I_{{\cal Q}}$, which for an empty \(AdS_3\) metric yields, i.e.,
\begin{eqnarray}
I_{bulk}=-\frac{1}{16\pi G}\int_{{\cal N}}{d^{3}x\sqrt{g}(R-2\Lambda)}-\frac{1}{8\pi G}\int_{\mathcal{M}}{d^{2}x\sqrt{\gamma}(K^{(\gamma)}-\Sigma^{(\gamma)})},
\end{eqnarray}
$\gamma$ and $\Sigma^{(\gamma)}$ are the induced metric and the surface tension on $\mathcal{M}$, respectively. $K^{(\gamma)}$ is the trace of the extrinsic curvature on the surface $\mathcal{M}$. On the other hand, for the boundary we have
\begin{eqnarray}
I_{{\cal Q}}=-\frac{1}{16\pi G}\int_{{\cal N}}{d^{3}x\sqrt{g}(R-2\Lambda)}-\frac{1}{8\pi G}\int_{{\cal Q}}{d^{2}x\sqrt{h}(K-\Sigma)},
\end{eqnarray}
where $h$ and $\Sigma$, are respectively the induced metric and the surface tension on ${\cal Q}$. With
\begin{eqnarray}
0\leq t\leq\tau,\quad
\epsilon\leq z\leq z_{\mathrm{IR}},\quad
-z\tan\theta_0\leq x\leq x_R .   
\end{eqnarray}
We can note that by localizing the surface $\mathcal{M}$ in $z=0$ we can write $\Sigma^{(\gamma)}=1/L$, $K^{\gamma}=2/L$ with $R=-6/L^{2}$, and removing the IR divergence by introducing a cutoff $\epsilon$ the $I_{bulk}$ becomes   
\begin{eqnarray}
I_{bulk}&=&\frac{\tau\,Lx_R}{4\pi\,G}\int^{z_{IR}}_{\epsilon}{\frac{dz}{z^{3}}}-\frac{1}{8\pi G}\int^{\tau}_{0}\int^{x_R}_{0}{d\tau dx\frac{L}{\epsilon^{2}}},\nonumber\\
        &=&-\frac{\tau\,Lx_R}{4\pi\,G}\left(\frac{1}{z^{2}_{IR}}-\frac{1}{\epsilon^{2}}\right)-\frac{\tau\,Lx_R}{4\pi\,G}\frac{1}{\epsilon^{2}},\nonumber
\end{eqnarray}
we obtain
\begin{equation}
I_{bulk}=-\frac{\tau\Delta\,x_RL}{4\pi\,z_{\rm IR}G}.
\end{equation}
This is only the bulk volume contribution. Gibbons--Hawking--York terms \cite{Takayanagi:2011zk,Fujita:2011fp}, the brane tension term, and holographic counterterms must be added to obtain the total renormalized gravitational action. For the boundary contribution, the brane ${\cal Q}$ is given by $x = -z\tan\theta_0$ and can be parametrized by $(t,z)$. Now, for the $I_{{\cal Q}}$ with $K=2\Sigma$, we have
\begin{eqnarray}
I_{{\cal Q}}&=&\frac{\tau\,Lx_R}{4\pi\,G}\int^{z_{max}}_{\epsilon}{\frac{dz}{z^{3}}}-\frac{1}{4\pi G}\int^{\tau}_{0}\int^{x_R}_{0}{d\tau dx\frac{L}{\epsilon^2}},\nonumber\\
        &=&-\frac{\tau\,Lx_R}{4\pi\,G}\left(\frac{1}{z^2_{max}}-\frac{1}{\epsilon^2}\right)-\frac{\tau\,Lx_R}{4\pi\,G}\frac{1}{\epsilon^2}=-\frac{\tau\,Lx_R}{4\pi\,G}\frac{1}{z^2_{max}},\nonumber
\end{eqnarray}
Then using the relation $I=I_{bulk}+I_{{\cal Q}}$, we have
\begin{eqnarray}
I=-\frac{\tau\Delta\,x_RL}{4\pi\,z_{\rm IR}G}-\tau\left[\frac{L}{4\pi\,z_{max} G}\tan\theta_0\right],
\end{eqnarray}

Now, using $z_{\rm IR} \to \infty$, we have

\begin{equation}
\boxed{I=\tau\left[\frac{L}{4\pi\,z_{max} G}\tan\theta_0\right].}
\end{equation}
Choosing $\tau=2\pi z_{\max}$ and using the partition function $Z=e^{-I}$, we obtain
\begin{equation}
{\cal S}_{bdry}=\frac{L}{2G}\tan\theta_0,\label{BDRY}
\end{equation}
The result above is a contribution from the regulated brane action, not the finite part of the complete Wheeler--DeWitt action. In particular,
choosing $\tau=2\pi z_{\max}$ does not turn the Lorentzian WDW action into the Euclidean partition function used to determine the
Affleck--Ludwig entropy. We therefore do not identify the expression obtained from $I_{{\cal Q}}$ with $S_{\rm bdry}=\log g_B$, nor do we
infer $\Delta\mathcal C_{\rm bdry}=\log g_B/(\pi\hbar)$ from it. Such an identification would require a comparison of complete WDW actions
with matched boundary data and a specified renormalization prescription. The calculation at this stage establishes sensitivity
to the brane embedding, but not a universal finite boundary complexity. We now proceed to the joint terms \cite{Shimaji:2018czt}.

\begin{figure}[!ht]
\begin{center}
\includegraphics[scale=0.75]{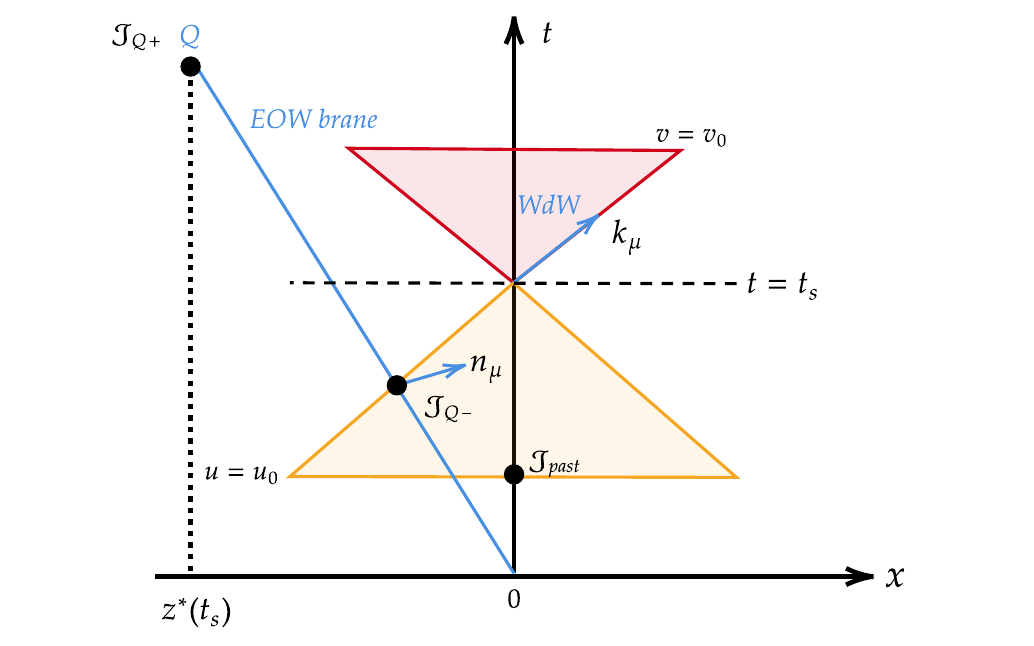}
\caption{WDW patch in Poincaré $AdS_3$. The null sheets (red: future, orange: past) originate at $t=t_s$ and bound the WDW patch. The EOW brane ${\cal Q}$ (blue) intersects the sheets at the joints $\mathcal{J}_{{\cal Q}\pm}$. The past vertex $\mathcal{J}_{\rm past}$ is where the two past sheets meet.The vector $n^\mu$ is the normal to the brane, and $k^\mu$ is the null generator of the future sheet.}\label{fig:WDW}
\end{center}
\end{figure}

The WDW patch is bounded by the null cones \(u=u_0\) and \(v=v_0\), where \(u=t-x\) and \(v=t+x\) are null coordinates in Poincaré \(AdS_3\) \cite{Dedushenko:2018aox}, see Fig.~\ref{fig:WDW}. In our case, the constant-time surface \(t=t_s\) is associated with two symmetric null sheets. In \((t,z)\) coordinates, the null generators are
\begin{equation}
  k^\mu = \frac{1}{L^2/z^2}\,\partial_\mu(t\mp z)\quad(\text{future/past}),
\end{equation}
normalized such that \(k^\mu k_\mu=0\). At the intersections \(\mathcal{J}_{Q\pm}\), the relevant vectors are the null generator \(k^\mu\) of the sheet and the normal \(n^\mu\) to the brane. The joint angle is
\begin{equation}\label{aJQ}
  a_{{\cal Q}_\pm} = \log\!\left|\frac{k\cdot n}{|k||n|}\right|.
\end{equation}
For the brane \eqref{braneEq}, with normal \eqref{normalQ} and null generator \(k^\mu\propto(1,\pm1,0)/z\) (temporarily working in units with \(L=1\)), we have
\begin{equation}
  k\cdot n = g_{\mu\nu}k^\mu n^\nu
  = \frac{1}{z^2}\left[\mp\cos\theta_0 + \sin\theta_0\right].
\end{equation}
The precise value depends on the orientation; for the future sheet (upper sign), one finds
\begin{equation}\label{kndot}
  (k\cdot n)_+ = \frac{\sin\theta_0 - \cos\theta_0}{z^2}.
\end{equation}
The area element at the joint is, for \(d=1+1\) and a point in the spatial direction, \(\sqrt{\sigma}=L/z\), corresponding to the transverse dimension of the brane at the intersection. Hence, the joint contribution at \(\mathcal{J}_{{\cal Q}_+}\) is
\begin{equation}\label{IjQ}
  I_{\rm joint,{\cal Q}} = \frac{1}{8\pi G}\int_{\mathcal{J}_{{\cal Q}_+}}\!\sqrt{\sigma}\,
    \log\!\left|\frac{k\cdot\bar k}{2}\right|\,d\theta,
\end{equation}
where \(\bar k\) denotes the generator of the oppositely directed null sheet (or the brane generator, depending on the geometry of the joint). At late times, \(t\gg L\), the joint \(\mathcal{J}_{{\cal Q}_+}\) occurs at \(z=z_*(t)\), where the null sheet intersects the brane. From the geometry,
\begin{equation}
  z_*(t) = \frac{t}{\tan\theta_0+1}\quad(\text{approximately}),
\end{equation}
and the joint angle is
\begin{equation}
  a_{{\cal Q}_+}(t) = \log\!\left(\frac{t}{L}\,\sin\theta_0\right) +\mathcal{O}(1).
\end{equation}

At the past vertex \(\mathcal{J}_{\rm past}\), two future-directed null generators \(k\) and \(\bar k\) meet. Their inner product is \(k\cdot\bar k=-2/z_{\rm past}^2\) (with \(L=1\)), so that
\begin{equation}\label{apast}
  a_{\rm past} = \log\!\left|\frac{k\cdot\bar k}{2}\right|
  = -\log(z_{\rm past}^2).
\end{equation}
Since \(z_{\rm past}\) is independent of \(T\), as it lies in the interior of \(AdS_3\), away from the brane, this contribution cancels in the renormalized difference. To ensure finiteness, we include the null counterterms introduced by Lehner \emph{et al.}~\cite{Lehner:2016vdi}, which guarantee invariance of the action under reparametrizations \(\lambda\to\tilde\lambda(\lambda)\) of the null generators. Defining \(\Theta=\partial_\lambda\log\sqrt{\gamma}\), namely, the expansion of the null generators, we have
\begin{equation}
  I_{\rm ct} = \frac{1}{8\pi G}\int_{\mathcal{N}}\!
    \sqrt{\gamma}\,\Theta\,\log(ct\,\Theta)\,d\lambda\,d\theta.
\end{equation}
For \(AdS_3\), the metric transverse to the null generator is trivial in \(d-1=1\) dimension:
\[
\gamma_{ab}=\gamma(\lambda)\delta_{ab},
\]
with \(\gamma=\gamma(\lambda)\) a scalar. The affine generators have \(\Theta=-1/\lambda\) in suitable coordinates, which gives
\begin{equation}
  I_{\rm ct} = \frac{1}{8\pi G}\int\!\frac{1}{\lambda}\,
    \log\!\left(\frac{ct}{\lambda}\right)\,d\lambda\,d\theta
  = -\frac{1}{8\pi G}\left[\log^2\!\lambda - \log(ct)\log\lambda\right]_{\lambda_1}^{\lambda_2}.
\end{equation}
The contribution to $\Delta I_{\rm ct}$ arising from the change in $T$ is given by
\begin{equation}\label{DeltaIct}
  \Delta I_{\rm ct} = \frac{1}{8\pi G}\,\Delta[\text{boundary terms in } \lambda],
\end{equation}
which is ultimately absorbed into the definition of $\Delta{\cal C}_{\rm metric}$. The joint contribution at the EOW brane, following \eqref{IjQ}, is
\begin{equation}\label{jointQexpl}
  I_{\rm joint}^{({\cal Q})}(T) = \frac{\sqrt{\sigma}}{8\pi G}\,a_{Q+}\big|_{z=z_*(t)}
  = \frac{L}{8\pi G\,z_*}\!\left[\log\!\left(\frac{t\sin\theta_0}{L}\right) + \mathcal{O}(1)\right].
\end{equation}
With $z_* = t/(1+\tan\theta_0)$, we have
\begin{equation}
  \frac{L}{z_*} = \frac{L(1+\tan\theta_0)}{t},
\end{equation}
and therefore
\begin{equation}\label{JointQT}
  I_{\rm joint}^{({\cal Q})}(T) = \frac{L(1+\tan\theta_0)}{8\pi G\,t}
    \log\!\left(\frac{t\sin\theta_0}{L}\right).
\end{equation}
For a comparison with a reference angle $\theta_{\rm ref}$, the expression in Eq.$\sim$(\ref{JointQT}) gives, within the approximations used to
obtain that equation,
\begin{eqnarray}
 \Delta I_{\rm joint}^{(Q)}(t)
\simeq \frac{L}{8\pi Gt}\Bigg[
 &(1+\tan\theta_0)
   \log\!\left(\frac{t\sin\theta_0}{L}\right)\\
 -&(1+\tan\theta_{\rm ref})
   \log\!\left(\frac{t\sin\theta_{\rm ref}}{L}\right)
\Bigg].   
\end{eqnarray}
Here $0<\theta_0,\theta_{\rm ref}<\pi/2$; the zero-tension configuration cannot be inserted directly into this approximate logarithmic expression without first treating its limit in the underlying joint calculation. In particular, $\Delta I_{\rm joint}^{(Q)}=O(\log t/t)$ and tends to zero in the formal large-$t$ limit. This term therefore cannot, by itself, yield a time-independent contribution proportional to $\log g_B$ \cite{Tang:2017xjc}.

The null-boundary counterterm has a distinct role: it removes the dependence of the action on the parametrization of the null generators. Its inclusion does not, on its own, fix a finite brane-localized counterterm or establish that its contribution cancels the joint difference. In particular, defining a finite
counterterm to be minus the expression above would make the cancellation true by construction, rather than derive it from a renormalization principle. We do not impose such a cancellation. The normalization of the null generators, the null-counterterm scale, the matching of cutoff surfaces, and any admissible finite
boundary terms must instead be specified before interpreting a finite relative WDW action. With these choices made explicit, the relative CA observable is
\begin{eqnarray}
\Delta\mathcal C_{\mathcal S}(T,T_{\rm ref})
=\frac{
 I_{\rm WDW}^{\rm ren}(T;\mathcal S)
-I_{\rm WDW}^{\rm ren}(T_{\rm ref};\mathcal S)}
{\pi\hbar},    
\end{eqnarray}
where $\mathcal S$ denotes the regulator-matching and renormalization prescription. The boundary entropy (\ref{BDRY}) was determined independently by the $I=I_{bulk}+I_{{\cal Q}}$ or the RT calculation as presented previously. One may organize a finite result in a specified scheme as
\begin{eqnarray}
\Delta\mathcal C_{\mathcal S}(T,T_{\rm ref})
=\frac{\log g_B(T)-\log g_B(T_{\rm ref})}{\pi\hbar}
+\mathcal R_{\mathcal S}(T,T_{\rm ref}),   
\end{eqnarray}
where $\mathcal R_{\mathcal S}$ collects the remaining geometric and scheme-dependent contributions. This decomposition is not a derivation that $\mathcal R_{\mathcal S}=0$. Establishing a universal equality with the boundary-entropy difference would require a calculation of the complete, regulator-matched WDW action and a demonstration that the remainder vanishes independently of admissible changes of scheme. This distinction is consistent with the prescription dependence found in \cite{Braccia:2019xxi}. The sewing analogy organizes the interface contributions, while the value of the relative CA observable remains a question for the full metric-dependent gravitational calculation.

\section{Thermal Extensions}\label{Sec4}

The interpretation of relative complexity as an interface quantity, established in the previous section, allows for natural extensions to thermal geometries, black holes, matter fields, defects localized on the EOW brane, and modified gravity theories \cite{Santos:2025fdp}. The objective of this section is to establish an organized prescription for a thermal geometry—specifically, to separate universal boundary contributions from local, scheme-dependent terms. We follow the principle of the Complexity-Action (CA) proposal, according to which
\begin{equation}
\mathcal{C}_{A}=\frac{I_{\rm WDW}}{\pi\hbar},
\label{eq:CAextension}
\end{equation}
where \(I_{\rm WDW}\) is the full gravitational action evaluated on the Wheeler–DeWitt (WDW) patch, including volume, boundary, and joint terms, as well as null counterterms, the brane contribution, and, when present, matter and defects \cite{Brown:2015bva,Brown:2015lvg,Lehner:2016vdi}.

The primary generalization relative to the vacuum case stems from the fact that the geometry now possesses a thermal scale, determined by the Hawking temperature \(T_{\rm H}\), and may include a region behind the horizon. Furthermore, the brane \(Q\) can support intrinsic fields and higher-codimension defects, simultaneously modifying its equilibrium condition and the localized terms of the action. The conceptual framework of TQFT remains valid for organizing the gluing procedure: the interface still carries state and boundary data. However, the weights assigned to these data are now determined by an effective action that is metric-dependent, thermal, and, in general, coupled to matter.

\subsection{Thermal Geometry and BTZ Black Hole with an EOW Brane}

We initially consider the planar BTZ black hole geometry,
\begin{equation}
ds^2=-f(r)dt^2+\frac{dr^2}{f(r)}+r^2dx^2,\,\, f(r)=\frac{r^2-r_h^2}{L^2},
\label{eq:BTZmetric}
\end{equation}
where \(L\) is the AdS radius and \(r_h\) is the horizon radius. The Hawking temperature is obtained by imposing regularity on the Euclidean section. For this purpose, we set \(t=-i\tau\), such that
\begin{equation}
ds_E^2=f(r)d\tau^2+\frac{dr^2}{f(r)}+r^2dx^2.
\label{eq:EuclideanBTZ}
\end{equation}
Near the horizon, we write
\begin{equation}
r=r_h+\delta r,\,\,f(r)=f'(r_h)\delta r+\mathcal{O}(\delta r^2).
\label{eq:nearhorizon}
\end{equation}
Since
\begin{equation}
f'(r)=\frac{2r}{L^2}, \qquad f'(r_h)=\frac{2r_h}{L^2},
\label{eq:fprime}
\end{equation}
the metric in the \((\tau,r)\) plane locally takes a regular polar form if \(\tau\) has the following period:
\begin{equation}
\beta=\frac{4\pi}{f'(r_h)}=\frac{2\pi L^2}{r_h}.
\label{eq:betaBTZ}
\end{equation}
Therefore,
\begin{equation}
T_{\rm H}=\beta^{-1}=\frac{r_h}{2\pi L^2}.
\label{eq:Hawkingtemperature}
\end{equation}
The physical region of \(AdS/BCFT\) is bounded by an EOW brane \({\cal Q}\). For a brane described implicitly by
\begin{equation}
F(r,x)=x-X(r)=0,
\label{eq:braneembedding}
\end{equation}
the unit normal is constructed from
\begin{equation}
n_{\mu}=\frac{\partial_\mu F}{\sqrt{g^{\alpha\beta}\partial_\alpha F\partial_\beta F}}.
\label{eq:braneNormal}
\end{equation}
As
\begin{equation}
\partial_\mu F=(0,-X'(r),1),
\label{eq:dF}
\end{equation}
we have
\begin{equation}
g^{\mu\nu}\partial_\mu F\partial_\nu F=f(r)X'(r)^2+\frac{1}{r^2}.
\label{eq:normdF}
\end{equation}
Thus,
\begin{equation}
n_\mu=\frac{(0,-X'(r),1)}{\sqrt{f(r)X'(r)^2+r^{-2}}}.
\label{eq:unitnormalBTZ}
\end{equation}
The minimum Euclidean gravitational action, including a brane with tension $T$, is
\begin{equation}
I_E=-\frac{1}{16\pi G}\int_{\cal N} d^3x\sqrt{g}\left(R-2\Lambda\right)-\frac{1}{8\pi G}\int_{\cal Q} d^2y\sqrt{h}\left(K-T\right) + I_{\rm ct}.
\label{eq:EuclideanActionBTZ}
\end{equation}
The bulk contribution between the horizon \(r_h\) and a UV cutoff \(r_c\) is
\begin{align}
I_{\rm bulk}^{(E)}&=-\frac{1}{16\pi G}\int_0^\beta d\tau\int dx\int_{r_h}^{r_c}dr\,r\left(-\frac{4}{L^2}\right)\nonumber\\
&=\frac{\beta}{4\pi G L^2}\int dx\int_{r_h}^{r_c}r\,dr\nonumber\\
&=\frac{\beta}{8\pi G L^2}\left(r_c^2-r_h^2\right)\int dx.
\label{eq:thermalbulk}
\end{align}
The spatial integral is to be understood over the region allowed by the presence of the brane. If \(x\) is compactified with length \(\ell_x\), then
\begin{equation}
\int dx=\ell_x.
\label{eq:spatiallength}
\end{equation}
In a BCFT on a half-line or an interval, the brane position effectively determines the integration domain and, consequently, the interface contribution to the free energy and the WDW action. The variational condition on the brane is
\begin{equation}
K_{ab}-(K-T)h_{ab}=0.
\label{eq:NeumannThermal}
\end{equation}
Since \({\cal Q}\) is two-dimensional, taking the trace of \eqref{eq:NeumannThermal} yields
\begin{equation}
K-2(K-T)=0,
\label{eq:traceNeumann}
\end{equation}
namely,
\begin{equation}
K=2T.
\label{eq:tracecondition}
\end{equation}
Replacing this result back into \eqref{eq:NeumannThermal}, we obtain
\begin{equation}
K_{ab}=Th_{ab}.
\label{eq:umbilicthermal}
\end{equation}
For the static embedding \(x=X(r)\), let
\[
D(r)=f(r)X'(r)^2+r^{-2},
\qquad
n_\mu=\frac{(0,-X'(r),1)}{\sqrt{D(r)}}.
\]
In coordinates \((t,r)\) on \(\mathcal Q\), the induced metric has
\[
h_{tt}=-f(r),\qquad
h_{rr}=f(r)^{-1}+r^2X'(r)^2.
\]
With the stated normal orientation, its time-time extrinsic-curvature component obeys
\[
\frac{K_{tt}}{h_{tt}}
=-\frac{f'(r)X'(r)}{2\sqrt{D(r)}}
=-\frac{rX'(r)}{L^2\sqrt{D(r)}} .
\]
Imposing \(K_{tt}=T h_{tt}\) and choosing the branch with \(X'(r)<0\) gives
\[
X'(r)^2
=\frac{T^2L^4}
 {r^2\left[r^2-T^2L^4 f(r)\right]}.
\]
Since \(f(r)=(r^2-r_h^2)/L^2\), the profile is
\[
X'(r)=
-\frac{TL^2}
 {r\sqrt{(1-T^2L^2)r^2+T^2L^2r_h^2}},
\qquad |LT|<1.
\]
Substitution into the radial component verifies \(K_{rr}=T h_{rr}\); thus the time-time equation is consistent with the full Neumann condition for this embedding. Reversing the normal reverses the corresponding sign convention for \(T\) and the selected branch of \(X'(r)\). Thus, the brane remains totally umbilical, but its embedding solution \(X(r)\) is deformed by the presence of the thermal factor \(f(r)\). In particular, the scale \(r_h\), which is absent in the Poincaré vacuum, allows the brane to interact geometrically with the horizon or terminate in a region behind it, depending on the boundary conditions and the holographic state considered.

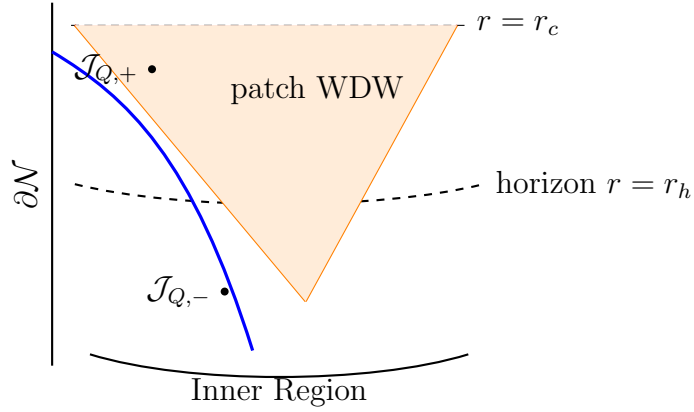
\begin{figure}[t]
\centering
\begin{tikzpicture}[scale=1.0,>=stealth]
  \draw[thick] (-3,-2.4) -- (-3,2.4);
  \node[rotate=90] at (-3.35,0) {$\partial {\cal N}$};

  \draw[thick,dashed] (-2.7,0) .. controls (-1.3,-0.35) and
  (1.3,-0.35) .. (2.7,0);
  \node[right] at (2.72,0) {horizon \(r=r_h\)};

  \draw[blue,very thick]
  (-3,1.75) .. controls (-2.0,1.15) and (-1.15,0.35) .. (-0.35,-2.2);
  \node[blue] at (-1.15,1.55) {\({\cal Q}\)};

  \draw[densely dashed] (-2.75,2.1) -- (2.45,2.1);
  \node[right] at (2.47,2.1) {\(r=r_c\)};

  \draw[thick] (-2.5,-2.25) .. controls (-1.2,-2.65) and
  (1.2,-2.65) .. (2.5,-2.25);
  \node at (0,-2.75) {Inner Region};

  \draw[red,thick] (-2.7,2.1) -- (0.35,0.35);
  \draw[red,thick] (2.35,2.1) -- (0.35,0.35);
  \draw[orange,thick] (-2.7,2.1) -- (0.35,-1.55);
  \draw[orange,thick] (2.35,2.1) -- (0.35,-1.55);

  \fill[red!15] (-2.7,2.1) -- (0.35,0.35) -- (2.35,2.1) -- cycle;
  \fill[orange!15] (-2.7,2.1) -- (0.35,-1.55) -- (2.35,2.1) -- cycle;

  \node at (0.5,1.25) {patch WDW};
  \fill (-1.68,1.52) circle (1.5pt);
  \node[left] at (-1.72,1.52) {\({\cal J}_{Q,+}\)};
  \fill (-0.72,-1.42) circle (1.5pt);
  \node[left] at (-0.76,-1.42) {\({\cal J}_{Q,-}\)};
\end{tikzpicture}
\caption{Qualitative diagram of a WDW patch in a thermal AdS$_3$/BCFT$_2$ geometry. The EOW brane \({\cal Q}\) modifies the physical domain and produces additional joints \({\cal J}_{{\cal Q},\pm}\) at the intersection with the null sheets.}
\label{fig:thermalWDW}
\end{figure}

To obtain an explicit result, it is necessary to specify the state and the physical region. We first consider the planar BTZ geometry without matter fields or defects, where \(x \sim x + \ell_x\), and the WDW patch of a two-sided thermofield double state anchored at symmetric boundary times \cite{Santos:2025fdp,Brown:2015lvg}. In the late-time regime, the CA action satisfies
$$
\frac{d I_{\rm WDW}}{dt} = 2M.
$$
The mass of the planar BTZ per coordinate length \(\ell_x\) is
$$
M = \frac{\ell_x r_h^2}{16\pi G L^2}.
$$
Consequently,
$$
\boxed{
\frac{d I_{\rm WDW}}{dt} = \frac{\ell_x r_h^2}{8\pi G L^2}
}
$$
and, according to the Complexity-Action prescription,
$$
\boxed{
\frac{d\mathcal C_A}{dt} = \frac{\ell_x r_h^2}{8\pi^2 G \hbar L^2}.
}
$$
In terms of the Hawking temperature, \(r_h = 2\pi L^2 T_{\rm H}\), which leads to
$$
\boxed{
\frac{d\mathcal C_A}{dt} = \frac{\ell_x L^2 T_{\rm H}^2}{2G\hbar}.
}
$$
Equivalently, the Bekenstein–Hawking entropy is given by \({\cal S}_{\rm BH} = \frac{\ell_x r_h}{4G}\), so that
$$
T_{\rm H} {\cal S}_{\rm BH} = \frac{\ell_x r_h^2}{8\pi G L^2} = 2M.
$$
Thus, the result can be expressed thermodynamically as
$$
\boxed{
\frac{d\mathcal C_A}{dt} = \frac{T_{\rm H} {\cal S}_{\rm BH}}{\pi\hbar} = \frac{2M}{\pi\hbar}.
}
$$
This represents the universal late-time growth of the CA complexity for the non-rotating BTZ black hole. In the BTZ background, the bulk action in a region \(\mathcal W\) of the WDW patch is
$$
I_{\rm bulk} = -\frac{1}{4\pi G L^2} \int_{\mathcal W} dt \, dr \, dx \, r.
$$
For an additional late-time slice of duration \(dt\), the relevant contribution comes from the interior region of the horizon. Integrating radially from \(r=0\) to \(r=r_h\), we find
$$
\frac{dI_{\rm bulk}}{dt} = -\frac{\ell_x}{4\pi G L^2} \int_0^{r_h} r \, dr = -\frac{\ell_x r_h^2}{8\pi G L^2},
$$
yielding
$$
\boxed{\frac{dI_{\rm bulk}}{dt} = -2M.}
$$
This is not yet the total result: the null joints near the singularity and at the meeting point of the null sheets provide the complementary contribution. At late times, the sum of joint terms, surface terms, and null counterterms yields
$$
\frac{dI_{\rm joint} + dI_{\rm nonnull} + dI_{\rm ct}^{\rm null}}{dt} = \frac{\ell_x r_h^2}{4\pi G L^2} = 4M.
$$
Therefore, \(\frac{dI_{\rm WDW}}{dt} = -2M + 4M = 2M\), as stated. In the presence of an EOW brane with tension \(T\), the correct expression is
$$
\frac{d\mathcal C_A}{dt} = \frac{1}{\pi\hbar} \left[ \frac{dI_{\rm bulk}}{dt} + \frac{dI_{\cal Q}}{dt} + \frac{dI_{\rm joint}}{dt} + \frac{dI_{\rm ct}^{\rm null}}{dt} \right].
$$
The equilibrium condition is \(K_{ab} = T h_{ab}\). This result explicitly shows the thermal deformation of the embedding due to the scale \(r_h\). In the limit \(r_h \to 0\), one recovers the behavior of a constant-tension brane in Poincaré AdS. The brane contribution \cite{Magan:2014dwa} to the growth is
$$
\frac{dI_{\cal Q}}{dt} = \frac{1}{8\pi G} \frac{d}{dt} \int_{{\cal Q}\cap\mathcal W} d^2y \, \sqrt{|h|} \, (K-T).
$$
Using the on-shell condition \(K=2T\), one obtains \(K-T=T\), and thus
$$
\boxed{
\frac{dI_{\cal Q}}{dt} = \frac{T}{8\pi G} \frac{d}{dt} \operatorname{Area}(\mathcal{Q} \cap \mathcal{W}) = \frac{L}{2G} \sinh^{-1}[\cot(\theta_0)].
}
$$
The correction due to the brane depends on the specific portion of \({\cal Q}\) contained within the WDW patch and the null joints \({\cal J}_{{\cal Q},\pm}\). Depending on whether the state is one-sided or two-sided, the time anchoring of the null sheets, and the physical domain selected by the brane, this contribution may vary. The most general form for the growth is
$$
\boxed{
\frac{d\mathcal C_A}{dt} = \frac{\ell_x r_h^2}{8\pi^2G\hbar L^2} + \frac{1}{\pi\hbar} \left[ \frac{dI_{\cal Q}}{dt} + \frac{dI_{{\cal J}_{{\cal Q},+}}}{dt} + \frac{dI_{{\cal J}_{{\cal Q},-}}}{dt} \right].
}
$$
For a comparison between two boundary conditions with the same temperature and induced metric at the cutoff, the renormalized complexity is given by
$$
\boxed{
\frac{d\mathcal C_A}{dt} = \frac{1}{\pi\hbar} \left[ \frac{\ell_x r_h^2}{8\pi G L^2} + \frac{L}{2G} \sinh^{-1}[\cot(\theta_0)] \right] = \frac{1}{\pi\hbar} [2M + {\cal S}_{bdry}].
}
$$
The product \(T_{\rm H}{\cal S}_{\rm BH} = 2M\) measures the energy scale of the thermal degrees of freedom driving the growth of the black hole interior. Conversely, \({\cal S}_{bdry}\) is the additional contribution determined by the interface—that is, by the holographic boundary condition implemented by the brane. Thus, complexity does not merely measure "how much thermal interior" the black hole possesses; it is also sensitive to how the system is terminated at the boundary, reflecting the interface structure of the \(BCFT\).

For a comparison between boundary conditions at fixed temperature and with matched induced data at the cutoff, the late-time growth must be obtained from the complete WDW action, including the brane, its null joints, and the chosen null-boundary counterterms. The reference BTZ contribution $2M$ does not, by itself, determine the brane-dependent correction. We therefore write schematically
\begin{equation}
\frac{d\mathcal C_A}{dt}
=\frac{1}{\pi\hbar}
\left[2M+\delta\dot I_{\rm interface}(\mathcal S)\right],   
\end{equation}
where $\delta\dot I_{\rm interface}(\mathcal S)$ denotes the net interface-associated correction in the specified setup and renormalization scheme. The present calculation does not establish that this correction equals the Affleck--Ludwig boundary entropy or that it is independent of the scheme \cite{Tang:2017xjc}.

\section{Conclusions and discussions}\label{Sec5}

In this work, we have utilized the gluing and cobordism axioms of Topological Quantum Field Theory to construct an organizing framework for boundary contributions in AdS/BCFT. By contrasting topological excision with the geometric partition of Lorentzian spacetimes, we clarified the operational distinction between fictitious internal boundaries—which vanish under standard variational cancellations—and physical End-of-the-World branes, which introduce localized variational, extrinsic curvature, and joint data \cite{Simon:2023hdq,Carqueville:2017fmn,Dedushenko:2018aox,Takayanagi:2011zk,Fujita:2011fp,Nishioka:2021cxe}.

Applying this framework to the Complexity equals Action (CA) proposal within the Wheeler–DeWitt patch, we demonstrated that holographic complexity exhibits an intrinsic sensitivity to the choice of boundary conditions. However, our analysis highlights crucial caveats regarding universality: the finite difference of the WDW action between distinct boundary configurations retains a non-trivial dependence on the embedding geometry and joint kinematics. Consequently, identifying this finite relative complexity directly with the universal Affleck–Ludwig boundary entropy ($\Delta\,C=log\,g_{{\cal B}}/\pi\hbar$) isolating a clean $S_{bray}$ shift in the late-time BTZ complexity growth, is strictly contingent upon a chosen renormalization scheme and boundary counterterm prescription \cite{Brown:2015lvg,Lehner:2016vdi,Santos:2025fdp,Braccia:2019xxi,Sato:2019kik,Flory:2017ftd,Heckman:2026beg,Yekta:2020wup,Babaei-Aghbolagh:2021ast,Couch:2021wsm,Fu:2018kcp}.

Rather than claiming a prescription-independent derivation of boundary complexity, our results establish the boundary data as an interface contribution that must be incorporated into the holographic dictionary of state preparation. Resolving whether an unambiguous, covariant counterterm prescription exists that isolates $logg_{{\cal B}}$  across arbitrary dimensional BCFTs and black hole backgrounds remains an essential open problem at the intersection of quantum information and gravitational sewing.

\acknowledgments
I would like to thank Marc Henneaux, Aldo Cotrone, Mario Flory, Jonathan Heckman, and Komeil Babaei Velni for the fruitful discussions. Fabiano F. Santos is partially supported by Conselho Nacional de Desenvolvimento Cient\'{\i}fico e Tecnol\'{o}gico (CNPq) under grant 302835/2024-5.

\end{document}